\documentclass[oldversion]{aa}
\usepackage{epsfig}
\usepackage{natbib}
\usepackage{lscape}
\usepackage{wasysym}
\bibpunct{(}{)}{;}{a}{}{,}
\begin{document}

\title{Chemical abundances in six nearby star-forming regions\thanks{Based on observations collected at Paranal Observatory, 
{\sc eso} (Chile) with the {\sc uves} and {\sc flames/uves} spectrographs at the VLT-Kueyen 
spectrograph (run IDs 075.C-0272(A) and 076.C-0524(A), respectively), with the {\sc feros} spectrograph at the {\sc eso/mpi} 
2.2-m telescope (program ID 070.C-0507(A)), as well as with the {\sc sophie} spectrograph at the {\sc ohp} observatory, France.}}

\subtitle{Implications for galactic evolution and planet searches around very young stars}

\author{
  N. C. Santos\inst{1} \and
  C. Melo\inst{2} \and
  D. J. James\inst{3} \and
  J. F. Gameiro\inst{1,4} \and
  J. Bouvier\inst{5}\and
  J. I. Gomes\inst{6}
}

\institute{
    Centro de Astrof{\'\i}sica, Universidade do Porto, Rua das Estrelas, 
    4150-762 Porto, Portugal
    \and
    European Southern Observatory (ESO), Casilla 19001, Santiago 19, Chile      
    \and
    Department of Physics \& Astronomy, Box 1807 Station B, Vanderbilt
    University, Nashville, TN 37235, USA
    \and
    Departamento de Matem\'atica Aplicada, Faculdade de Ci\^encias da 
    Universidade do Porto, Portugal
    \and
    Laboratoire d'Astrophysique, Observatoire de Grenoble, BP-53, 
    38041 Grenoble, France
    \and
    Centro de Astrof{\'\i}sica da Universidade de Lisboa, Tapada 
    da Ajuda, 1349-018 Lisboa, Portugal
}

\date{Received 16/11/2007; accepted 16/01/2008}

\abstract{
In this paper we present a study of chemical abundances in six 
star-forming regions. Stellar parameters and metallicities are derived 
using high-resolution, high S/N spectra of weak-line T-Tauri 
stars in each region. The results show that nearby star-forming 
regions have a very small abundance dispersion (only 0.033\,dex in [Fe/H]).
The average metallicity found is slightly below that of the Sun,
although compatible with solar once the errors are taken
into account. The derived abundances for Si and Ni show that the
observed stars have the abundances typical of Galactic thin disk 
stars of the same metallicity. The impact of these observations 
is briefly discussed in the context of the Galactic chemical 
evolution, local inter-stellar medium abundances, and in
the origin of metal-rich stars in the solar neighbourhood 
(namely, stars more likely to harbour planets). The implication 
for future planet-search programmes around very young, nearby 
stars is also discussed.
  }

\authorrunning{Santos et al.}
\maketitle

\section{Introduction}

As of November 2007, about 250 extra-solar planets have been 
discovered orbiting solar-type stars\footnote{For a continuously 
updated list see tables at http://www.exoplanets.eu or 
http://www.exoplanet.eu}, most of which were detected thanks 
to the development of high precision radial-velocity 
instruments \citep[for a review see][]{Udry-2007}.  Complementary 
high accuracy spectroscopic studies have shown that stars 
hosting giant  planets are particularly metal-rich when 
compared with single field stars \citep[e.g.][]{Gonzalez-1997,
Gonzalez-2001,Santos-2001,Santos-2004b,Santos-2005a,
Fischer-2005}. This observation is helping to better 
understand the processes of planet formation and 
evolution.

The high metal content of some of the planet-hosts has 
raised a number of interesting questions regarding their 
origin. It has been proposed that the high metallicities 
could be the result of the in-fall of planetary (metal-rich) 
material into the stellar outer convective envelope 
\citep[e.g.][]{Gonzalez-1998,Murray-1998}, although current 
results do not support this hypothesis
\citep[][]{Santos-2004b,Santos-2005a,Fischer-2005}. The 
excess metallicity observed likely reflects the higher 
(average) metal abundances of the clouds of gas and dust 
that originated the star and planet systems. Such 
conclusions are also supported by the most recent planet 
formation models \citep[][]{Ida-2004b,Benz-2006}.

Because the large majority of planet hosting stars are 
solar-neighbourhood thin-disk objects \citep[e.g.][]{Ecuvillon-2007}, 
one would expect nearby T-Tauri stars and star-forming
regions [{\sc sfr}s] that are metal-rich to exist. 
Their existence is expected if a well-defined
age-metallicity relation exists in the solar-neighbourhood 
\citep[][]{Pont-2004b,Haywood-2006}. Previous studies 
\citep[][]{Padgett-1996,James-2006}, however, have not succeeded 
in finding evidence of high metallicity, 
T-Tauri stars in nearby {\sc sfr}s. Whether this is due to the absence of 
such regions, or to the relatively small number of stars 
and {\sc sfr}s observed was not clear. 

Considering the strong metallicity-giant planet connection, 
the detection of nearby, metal-rich {\sc sfr}s would provide 
preferential targets for future planet searches around young 
T-Tauri stars. The detection of planets orbiting T-Tauri 
stars would provide observational constraints of paramount 
importance for the study of planet formation scenarios, 
since it gives us some hint about the timescales for planetary 
formation.

In this paper, we continue the metallicity programme started 
in \citet[][]{James-2006}, by increasing the number of observed 
young T-Tauri stars in each {\sc sfr}, and by adding abundance 
determinations for objects that have not yet been studied. 
In all, we present accurate stellar parameters and chemical 
abundances for 19 weak line T-Tauri stars [{\sc wtts}s] in six 
nearby {\sc sfr}s, namely Chamaeleon, Corona Australis, Lupus, 
Rho-Ophiuchus, Taurus and the Orion Nebula Cloud. The results 
of this study are presented and their implications discussed.

\section{The sample}\label{sec:sample}

For {\sc wtts}s in the Chamaeleon, Corona Australis, Lupus, 
Rho-Ophiuchus, and Taurus {\sc sfr}s, the sample used in 
the current paper to derive stellar parameters and chemical
abundances was chosen in an identical manner to that in 
\citet[][]{James-2006}, to which we refer the reader for 
extensive details. In brief, {\sc wtts} catalogues for each 
{\sc sfr} were constructed from {\sc rosat} All-Sky Survey 
[{\sc rass}] detections in and around each {\sc sfr}. We then 
selected those stars having spectral types between G and 
early-K and visual magnitudes of 12 or brighter, which also 
have weak signatures of infrared excess and magnetic 
activity ({\em e.g.,} having H$\alpha$ equivalent widths $<<$10\r{A}). 
We also restricted our sample to objects exhibiting substantial 
equivalent widths [{\sc ew}s] of the lithium resonance line at 
6708\r{A} (whose substantial abundance is indicative of youth), 
as well as those stars which are not components of multiple 
systems, as judged from existing kinematic data. 

In \citet[][]{James-2006}, we presented the results of the 
analysis of a sub-sample of these stars, whose spectra 
were obtained using the {\sc feros} spectrograph 
(2.2-m {\sc eso}/{\sc mpi} telescope). The data were used 
to study the chemical abundances of {\sc wtts}s in 3 southern 
star-forming regions (Chamaeleon, CrA and Lupus). 
Unfortunately, the small aperture of the telescope did not 
allow us to obtain high S/N data for that many interesting 
targets. The number of stars observed was thus quite low, 
and the S/N of their spectra was often not of a quality 
sufficient to derive accurate stellar parameters and 
abundances. The use of a high resolution spectrograph 
installed on a larger telescope was thus mandatory to 
improve our results. In this paper, we present an 
extension of the study initiated in \citet[][]{James-2006}, 
using new high S/N, high resolution spectroscopic 
observations, predominantly obtained using 8m-class 
telescopes.

\begin{figure}[t]
\resizebox{\hsize}{!}{\includegraphics{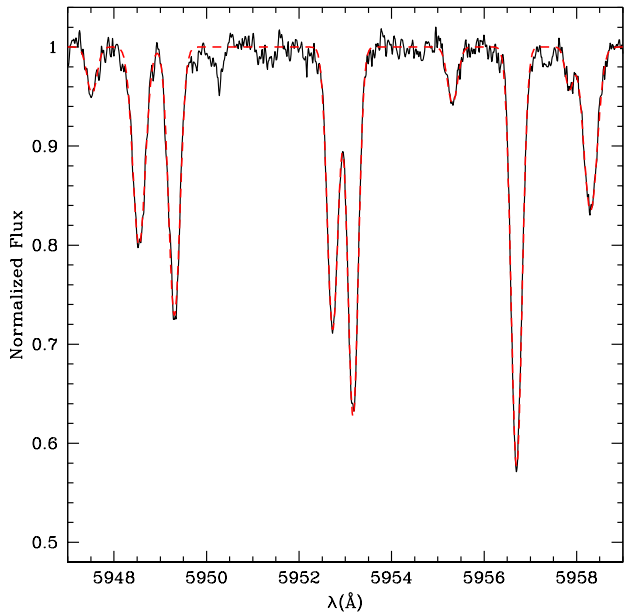}}
\caption{Sample UVES spectrum for one of the stars in our sample (RX\,J1159.7$-$7601) in the
region around 5950\AA. The spectrum presented has a S/N ratio of $\sim$140. 
Gaussian fits to most of the lines in the 
spectral domain presented are shown (dashed line). The EWs of the lines 
go from 12 to 130 m\AA, the typical values for the \ion{Fe}{i} lines in WTTS 
spectra used to derive stellar parameters and metallicity.} 
\label{fig:spectrum}
\end{figure}

After our new stellar spectra were reduced 
(see Sect.\,\ref{sec:observations}) we restricted a detailed 
spectroscopic analysis to narrow-lined stars, {\em i.e.,} 
those having low projected equatorial rotational velocity 
($<\sim15$km\,s$^{-1}$). Calculations of projected equatorial 
rotational velocity [$v\,\sin{i}$] for all stars were derived 
using the width of the cross-correlation function, using 
a method described by \citet[][]{Melo-2001} 
\citep[see also][]{James-2006}. This is imperative because 
higher rotational velocities pose important limitations 
on the accuracy of derived stellar parameters and chemical 
abundances (see Sect.\,\ref{sec:analysis}). 

\begin{table*}
\caption[]{List of WTTS observed in each {\sc sfr} and derived stellar parameters, metallicities, and
estimated $v\,\sin{i}$ values. The number of Fe~{\sc i} and Fe~{\sc ii} lines used 
for the spectroscopic analysis is also listed, together 
with the {\sc rms} around the average abundance for each set 
of lines. Stellar parameters and [Fe/H] derived from {\sc feros} 
spectra ([4]) were taken from \citet[][]{James-2006}.}
\begin{tabular}{lcccrcccc}
\hline
\hline
Star    & T$_{\mathrm{eff}}$ & $\log{g}_{spec}$ & $\xi_{\mathrm{t}}$ & \multicolumn{1}{c}{[Fe/H]} & Inst$^\dagger$ & N          & $\sigma$      & $v\,\sin{i}$  \\       
        & [K]                &  [cm\,s$^{-2}$]  &  [km\,s$^{-1}$]    &        &        & (Fe~{\sc i,ii})       &(Fe~{\sc i, ii})   & [km\,s$^{-1}$] \\
\hline
\multicolumn{9}{l}{Cha} \\
RX J1140.3-8321  & 4664$\pm$~94   & 4.17$\pm$0.52    & 2.11$\pm$0.15    & -0.22$\pm$0.11      & [1]  & 31,4           & 0.11,0.29 & 11      \\
RX J1140.3-8321  & 4814$\pm$154  & 4.07$\pm$0.92    & 2.15$\pm$0.23    & -0.14$\pm$0.15      & [4] &                 &           &          \\
RX J1158.5-7754A & 4774$\pm$101  & 4.08$\pm$0.41    & 2.11$\pm$0.15    & -0.28$\pm$0.11      & [1]  & 29,5           & 0.10,0.21 & 11       \\
RX J1158.5-7754A & 4810$\pm$142  & 3.87$\pm$0.51    & 2.11$\pm$0.22    & -0.26$\pm$0.15      & [4] &                 &           &          \\
RX J1159.7-7601  & 4740$\pm$~81   & 4.37$\pm$0.33    & 1.70$\pm$0.12    & -0.08$\pm$0.08      & [1]  & 32,5           & 0.08,0.17 & 9       \\
RX J1159.7-7601  & 4836$\pm$146  & 4.23$\pm$0.33    & 2.07$\pm$0.22    & -0.18$\pm$0.16      & [4] &                 &           &         \\
RX J1233.5-7523  & 5524$\pm$~53   & 4.49$\pm$0.16    & 1.22$\pm$0.07    & -0.03$\pm$0.06      & [1]  & 35,11          & 0.06,0.07 & 6       \\
RX J1233.5-7523  & 5469$\pm$~62   & 4.40$\pm$0.22    & 1.52$\pm$0.10    & -0.09$\pm$0.08      & [4] &                 &           &         \\
\hline
\multicolumn{9}{l}{CrA} \\
CrAPMS2         & 5400$\pm$164    & 4.06$\pm$0.45    & 2.41$\pm$0.30    & -0.09$\pm$0.18    & [4] &                 &           &         \\
CrAPMS-4SE      & 5236$\pm$~58    & 4.33$\pm$0.17    & 2.15$\pm$0.09    & -0.02$\pm$0.08    & [1]  & 35,8           & 0.07,0.08 & 10      \\
CrAPMS-4SE      & 5335$\pm$101    & 4.59$\pm$0.41    & 2.24$\pm$0.17    & -0.02$\pm$0.12    & [4] &                 &           &         \\
RX J1839.0-3726  & 5150$\pm$~66    & 4.40$\pm$0.22    & 2.45$\pm$0.11    & -0.10$\pm$0.08    & [1]  & 36,9           & 0.07,0.11 & 10     \\
RX J1839.0-3726  & 5293$\pm$127    & 4.51$\pm$0.42    & 2.19$\pm$0.21    & -0.01$\pm$0.15    & [4] &                 &           &        \\
\hline
\multicolumn{9}{l}{Lup} \\
RX J1507.2-3505  & 5169$\pm$~63    & 4.31$\pm$0.33    & 2.09$\pm$0.09    & -0.06$\pm$0.08    & [1]  & 37,9           & 0.07,0.17  & 13     \\
RX J1507.2-3505  & 5155$\pm$148    & 4.37$\pm$0.49    & 2.53$\pm$0.26    & -0.08$\pm$0.14    & [4] &                 &            &        \\
RX J1546.6-3618  & 5223$\pm$~83    & 4.51$\pm$0.22    & 2.22$\pm$0.12    & -0.04$\pm$0.09    & [1]  & 34,8           & 0.09,0.10  & 6     \\
RX J1546.6-3618  & 5062$\pm$109    & 4.09$\pm$0.53    & 2.12$\pm$0.16    & -0.12$\pm$0.14    & [4] &                 &            &        \\
RX J1547.6-4018  & 5006$\pm$~61    & 4.29$\pm$0.24    & 1.87$\pm$0.08    & -0.06$\pm$0.07    & [1]  & 34,7           & 0.07,0.12  & 12    \\
RX J1547.6-4018  & 5045$\pm$120    & 4.18$\pm$0.61    & 2.23$\pm$0.20    & -0.14$\pm$0.13    & [4] &                 &            &       \\
RX J1601.1-3320  & 5479$\pm$152    & 3.70$\pm$0.35    & 2.49$\pm$0.26    & -0.05$\pm$0.18    & [4] &                 &            &       \\
RX J1523.4-4055  & 4923$\pm$~87    & 4.74$\pm$0.37    & 2.05$\pm$0.16    & -0.03$\pm$0.09    & [1]  & 34,6           & 0.09,0.19  & 10    \\
\hline
\multicolumn{9}{l}{Rho-Ohp} \\
RX J1620.7-2348  & 4867$\pm$107    & 4.40$\pm$0.29    & 2.33$\pm$0.17    & -0.08$\pm$0.12    & [1]  & 35,7           & 0.12,0.13  & 9     \\
\hline
\multicolumn{9}{l}{ONC} \\
JW157           & 4877$\pm$~83      & 3.57$\pm$0.28    & 1.97$\pm$0.09    & -0.06$\pm$0.11   & [2]& 36,10            & 0.10,0.14  & 7      \\
JW589           & 4927$\pm$108     & 3.90$\pm$0.30    & 1.95$\pm$0.13    & -0.12$\pm$0.12   & [2]& 33,5             & 0.11,0.15  & 14     \\
JW706           & 4599$\pm$~79      & 3.86$\pm$0.45    & 1.50$\pm$0.10    & -0.17$\pm$0.09   & [2]& 30,4             & 0.09,0.25  & 12     \\
\hline
\multicolumn{9}{l}{Tau} \\
1RXSJ052146.7   & 4921$\pm$~59      & 4.05$\pm$0.29    & 1.92$\pm$0.07    & -0.07$\pm$0.07   & [3]& 34,5             & 0.07,0.15  & 13      \\
1RXSJ053650.0   & 4918$\pm$112     & 4.47$\pm$0.29    & 1.96$\pm$0.17    & -0.18$\pm$0.10   & [3]& 28,3             & 0.09,0.12  & 14      \\
1RXSJ053931.0   & 5430$\pm$~53      & 4.09$\pm$0.23    & 1.87$\pm$0.08    &  0.05$\pm$0.07   & [3]& 36,12            & 0.06,0.10  & 11      \\
\hline
\end{tabular}
\\ $\dagger$ Instruments used: [1] {\sc uves}; [2] {\sc flames}; [3] {\sc sophie}; [4] {\sc feros} data from \citet[][]{James-2006}
\label{table:parameters}
\end{table*}

One-dimensional kinematic data (radial velocities [{\sc rv}s]) 
of the target stars were compared with the expected systemic  
velocities of their parent {\sc sfr} (using values calculated 
and compared in James et al. 2006). Target stars were only 
considered for a detailed spectroscopic abundance analysis 
if their {\sc rv}s were invariant ({\i.e.,} not in a binary 
system) and they matched that of their natal cloud. A full 
description of the membership, mass and age properties of 
our target stars, as well as details on the rapid rotators, 
is presented in James et al. (in prep.). 

For stars in the Orion Nebula Cloud, data and membership 
information were taken from the studies of 
\citet[][]{Sicilia-Aguilar-2005}, \citet[][]{Rhode-2001}, and 
\citet[][]{Jones-1988}.  

We note that one star from the 
Chamaeleon {\sc sfr} (RX J1201.7-7959) whose metallicity 
was derived in \citet[][]{James-2006} is now excluded 
from the sample, because its {\sc rv}s indicate 
that it is a spectroscopic binary system.

\section{Observations}\label{sec:observations}

Between April and September 2005, we obtained a series 
of high resolution spectra for {\sc wtts}s in 4 young 
{\sc sfr}s (Chamaeleon, Corona Australis, Lupus and 
Rho-Ophiuchus) using the {\sc uves} spectrograph on 
the {\sc vlt}'s Kueyen 8.2-m telescope ({\sc eso} 
program ID 075.C-0272). The spectra were obtained 
in Service Mode using the Red\,580 mode, and cover 
the wavelength domain between 4800 and 7000\r{A}, 
with a gap between 5700 and 5850\r{A}. A binning of 
2x2 pixels and a slit width of 0.9 arcsec was used, 
leading to a resolution of R=$\lambda/\Delta\lambda\sim$50\,000. 
A sample UVES spectrum used is presented in Fig.\,\ref{fig:spectrum}.

\begin{table*}
\caption[]{Abundances of Si and Ni for the stars in Table\,\ref{table:parameters} that were 
observed with {\sc uves}, {\sc flames} or {\sc sophie} 
spectrographs. The [Fe/H] values listed in 
Table\,\ref{table:parameters} are also presented 
for comparison. }
\begin{tabular}{lccccccc}
\hline
\hline
Star    & [Fe/H] & [Si/H] & $\sigma$([Si/H]) & N(Si) & [Ni/H] & $\sigma$([Ni/H]) & N(Ni)\\       
\hline
\multicolumn{8}{l}{Cha} \\
RX J1140.3-8321  &-0.22 &  -0.19 &  0.08   &   4    &   -0.30  & 0.13    & 15    \\ 
RX J1158.5-7754A &-0.28 &  -0.27 &  0.07   &   4    &   -0.34  & 0.14    & 14    \\ 
RX J1159.7-7601  &-0.08 &  -0.05 &  0.07   &   5    &   -0.13  & 0.09    & 18    \\
RX J1233.5-7523  &-0.03 &   0.00 &  0.04   &   9    &   -0.07  & 0.04    & 18    \\
\hline
\multicolumn{8}{l}{CrA} \\
CrAPMS-4SE      &-0.02 &  -0.01 &  0.07   &   9    &   -0.08  & 0.13    & 18    \\ 
RX J1839.0-3726  &-0.10 &  -0.10 &  0.10   &   8    &   -0.22  & 0.11    & 16    \\ 
\hline
\multicolumn{8}{l}{Lup} \\
RX J1507.2-3505  &-0.06 &  -0.02 &  0.09   &   8    &   -0.18  & 0.08    & 14    \\ 
RX J1523.4-4055  &-0.03 &  -0.01 &  0.10   &   6    &   -0.13  & 0.11    & 14    \\ 
RX J1546.6-3618  &-0.04 &  -0.05 &  0.07   &   9    &   -0.11  & 0.10    & 18    \\ 
RX J1547.6-4018  &-0.06 &  -0.03 &  0.09   &   8    &   -0.15  & 0.12    & 12    \\ 
\hline
\multicolumn{8}{l}{Rho-Oph} \\
RX J1620.7-2348  &-0.08 &  -0.04 &  0.16   &   6    &   -0.12  & 0.12    & 17    \\ 
\hline
\multicolumn{8}{l}{ONC} \\
JW147           &-0.06 &  -0.07 &  0.12   &   9    &   -0.18  & 0.07    & 21    \\ 
JW589           &-0.12 &  -0.20 &  0.12   &   4    &   -0.18  & 0.09    & 11    \\ 
JW706           &-0.17 &  -0.07 &  0.13   &   6    &   -0.16  & 0.08    & 13    \\ 
\hline
\multicolumn{8}{l}{Tau} \\
1RXSJ052146.7   &-0.07 &   0.01 &  0.08   &   7    &   -0.12  & 0.13    & 11    \\ 
1RXSJ053650.0   &-0.18 &  -0.14 &  0.11   &   4    &   -0.23  & 0.13    & 11    \\ 
1RXSJ053931.0   & 0.05 &   0.05 &  0.09   &  10    &   -0.03  & 0.09    & 13    \\
\hline
\end{tabular}
\label{table:elements}
\end{table*}

To complement the {\sc uves} observations, six {\sc wtts}s 
in two more star-forming regions were observed. 
Observations of objects in the Orion Nebula Cloud 
[{\sc onc}] were carried out using the fibre link of the 
{\sc flames}-{\sc uves} spectrograph ({\sc eso} program 
ID 076.C-0524). The observations were carried out in 
Service Mode between October 2005 and March 2006. The 
{\sc ccd} was used in a 1x1 binning mode, and the resultant 
spectra have a resolution of $\sim$50\,000 and cover the 
same spectral domain as the {\sc uves} data. Finally, 
spectra for {\sc wtts} in the northern Taurus {\sc sfr} 
were obtained in December 2006 using the {\sc sophie} 
fibre-fed spectrograph at the 1.93-m telescope of the 
Observatoire de Haute Provence [{\sc ohp}], France. The 
spectra were obtained in the High-Efficiency mode 
(R$\sim$40\,000), positioning one of the fibres on the 
object and the other on the sky. Background light 
subtraction was performed using the sky ``spectrum'', 
since the instrument data reduction software does 
not allow for the subtraction of inter-order scattered 
light.

For all data sets, spectral extraction and data 
reduction was achieved using the available instrument 
pipelines supplied by the observatories. Wavelength 
calibration was done using the spectrum of a 
thorium-argon lamp. The spectra were then Doppler 
corrected and set at a zero velocity. When more than 
one spectrum of the same star was available, they were 
co-added. Final spectra have S/N ratios between 
100 and 200, and are of sufficient quality to derive 
stellar parameters and chemical abundances with a 
reasonable accuracy. In Table\,\ref{table:parameters}, 
we list fundamental physical properties for all the 
observed stars in each target {\sc sfr}. 

\section{Spectroscopic analysis}\label{sec:analysis}

\subsection{Stellar parameters and chemical abundances}

As in \citet[][]{James-2006}, stellar parameters and 
iron abundances for the targets were derived in 
{\sc lte}, using a grid of plane-parallel, {\sc atlas}9 
model atmospheres \citep[][]{Kurucz-1993} and the 2002 
version of the radiative transfer code {\sc moog} 
\citep[][]{Sneden-1973}. The methodology used is 
described in detail in \citet[][ and references 
therein]{Santos-2004b}. The full spectroscopic 
analysis is based on the {\sc ew}s of a set of Fe~{\sc i} 
and Fe~{\sc ii} weak lines, by imposing ionization 
and excitation equilibrium, as well as a zero 
slope between the abundances given by individual 
lines and their equivalent width. The errors 
in the stellar parameters were derived using 
the same methodology as described in \citet[][]{Gonzalez-1998}.

For FGK dwarfs, the stellar parameters obtained 
using this methodology were shown to be compatible 
with other estimates in the literature. In particular, 
the effective temperatures derived are very close to 
the ones obtained using recent applications of the 
infra-red flux method \citep[e.g.][]{Casagrande-2006}.

The results of these analysis are presented in 
Table\,\ref{table:parameters}, where we also compare 
the values derived using {\sc uves} spectra for 
stars in common with the \citet[][]{James-2006} sample. 
We note that in their paper, James et al. used 
lower quality spectra taken with {\sc feros}. A comparison 
of their results and our new ones shows that besides 
the larger error bars, the two datasets are in close 
agreement (well within error bars). For the duration 
of this paper, we will only consider the higher 
quality values derived from {\sc uves} spectra, for 
those stars having both {\sc feros} and {\sc uves} 
measurements.

\begin{table*}
\caption[]{Weighted average and dispersion of the metallicities and [X/Fe] (X=Si, Ni) abundance 
ratios derived for each {\sc sfr}. N and N' are the number of stars used for the determination of $<$[Fe/H]$>$ 
and $<$[X/H]$>$, respectively. For Rho-Oph, since only one star was measured, the {\sc rms} of the values 
correspond to the rms of the abundance determination for the star.}
\begin{tabular}{lcccccccc}
\hline
\hline
SFR      & $<$[Fe/H]$>$ & rms & N & $<$[Si/Fe]$>$ & $\sigma$([Si/Fe]) & $<$[Ni/Fe]$>$ & $\sigma$([Ni/Fe]) & N'\\
\hline
Cha      & $-$0.11 & 0.11 & 4 & 0.03 &  0.01  & $-$0.05  & 0.02   & 4\\
CrA      & $-$0.06 & 0.05 & 3 & 0.01 &  0.01  & $-$0.09  & 0.04   & 2\\
Lup      & $-$0.05 & 0.01 & 5 & 0.02 &  0.02  & $-$0.10  & 0.02   & 4\\
Rho-Oph  & $-$0.08 & 0.12 & 1 & 0.04 &  0.16  & $-$0.04  & 0.12   & 1\\
ONC      & $-$0.13 & 0.06 & 3 & 0.00 &  0.09  & $-$0.06  & 0.07   & 3\\
Tau      & $-$0.05 & 0.11 & 3 & 0.04 &  0.04  & $-$0.07  & 0.02   & 3\\
\hline
\end{tabular}
\label{table:average}
\end{table*}

As was discovered in the studies by \citet[][]{James-2006} and 
\citep[][]{Padgett-1996}, the derived micro-turbulence velocities 
are considerably above the ones found for main-sequence dwarfs 
with similar temperature detailed in the literature. The cause 
of these discrepancies is not yet clear to us, although the 
high values may be reflecting the effect of magnetic fields 
\citep[][]{Steenbock-1981}. 

Abundances for silicon (alpha-element) and nickel (iron-peak 
element) were derived using the methodology and atomic line 
lists described in \citet[][]{Gilli-2006} and \citet[][]{Santos-2006b}, 
and are listed in Table\,\ref{table:elements}. Analysis of these 
two elements was performed because they have several well 
defined lines in the spectra of solar type stars, and because 
they do not present any strong dependence of derived abundance 
with effective temperature \citep[][]{Bodaghee-2003,Gilli-2006}. 
This increases our confidence of uniformity in the comparison 
with abundances derived for solar-type field stars, presented 
in \citet[][]{Gilli-2006}.

\subsection{Veiling}

The spectroscopic analysis described above relies on the 
measurement of {\sc ew}s for a set of Fe~{\sc i} and Fe~{\sc ii} 
lines. This means that any contribution of veiling, related 
to stellar accretion phenomena, could lead to erroneous values 
for the stellar parameters and effective temperatures. Although 
high levels of veiling are not expected for the {\sc wtts} in our 
sample \citep[][]{Hartigan-1995}, we err on the side of caution, 
and have performed a more detailed analysis. 

\begin{table}
\caption[]{Veiling determinations for the stars listed in Table\,\ref{table:parameters} that were 
observed with {\sc uves}, {\sc flames} or {\sc sophie} spectrographs.}
\begin{tabular}{lcc}
\hline
\hline
Star  &  Veil &  Veil \\       
      & $<$5700\r{A} & $>$5850\r{A} \\
\hline
\multicolumn{3}{l}{Cha} \\
RX J1140.3-8321    & -0.01$\pm$0.04  & -0.07$\pm$0.06 \\
RX J1158.5-7754A  & 0.03$\pm$0.04   & 0.00$\pm$0.05 \\
RX J1159.7-7601    & -0.01$\pm$0.03  & -0.05$\pm$0.06 \\
RX J1233.5-7523    & 0.02$\pm$0.01   & 0.04$\pm$0.02 \\
\hline
\multicolumn{3}{l}{CrA} \\
CrAPMS-4SE       & 0.00$\pm$0.04   & -0.04$\pm$0.04  \\
RX J1839.0-3726   & 0.00$\pm$0.05   & -0.05$\pm$0.05  \\
\hline
\multicolumn{3}{l}{Lup} \\
RX J1507.2-3505  & 0.00$\pm$0.05    & -0.03$\pm$0.05 \\
RX J1546.6-3618  & -0.01$\pm$0.04   & -0.05$\pm$0.05 \\
RX J1547.6-4018  & 0.00$\pm$0.03    & -0.01$\pm$0.04 \\
RX J1523.4-4055  & -0.02$\pm$0.06   & -0.07$\pm$0.07 \\
\hline
\multicolumn{3}{l}{Rho-Ohp} \\
RX J1620.7-2348  & 0.01$\pm$0.04    & -0.04$\pm$0.06 \\
\hline
\multicolumn{3}{l}{ONC} \\
JW157           & 0.00$\pm$0.05   & -0.06$\pm$0.08 \\
JW589           & 0.01$\pm$0.10   & -0.04$\pm$0.08 \\
JW706           & 0.02$\pm$0.03   & -0.03$\pm$0.06 \\
\hline
\multicolumn{3}{l}{Tau} \\
1RXSJ052146.7   & 0.00$\pm$0.04   & -0.03$\pm$0.06 \\
1RXSJ053650.0   & 0.00$\pm$0.03   & -0.02$\pm$0.05 \\
1RXSJ053931.0   & 0.04$\pm$0.05   &  0.00$\pm$0.05 \\
\hline
\end{tabular}
\label{table:veiling}
\end{table}

The veiling (defined as the ratio between the excess 
continuum flux and the photospheric continuum contribution 
to the spectrum) was derived using a method based on 
that described in \citet[][]{Hartigan-1989}. With this method, 
the line depth of photospheric absorption lines are compared 
with a reference spectrum within a small wavelength band 
(a few tens of {\r{A} wide), and it is assumed that the 
veiling be constant in this region. The wavelength coverage 
of our spectra is almost complete from 4800 to 7000\r{A} 
with a small gap between 5700 and 5850\r{A}. For each object, 
a reference star (with similar temperature as the object) 
was rotationally broadened to match the $v\,\sin{i}$ of the 
object, and the veiling computed in intervals of 20\r{A}. 
The reference stars are dwarfs of similar effective 
temperature and metallicity to the targets, and were 
the ones used in previous chemical abundances studies 
of stars with and without extra-solar planets 
\citep[][]{Santos-2004b,Santos-2005a,Sousa-2006}.

\begin{figure}[t]
\resizebox{\hsize}{!}{\includegraphics{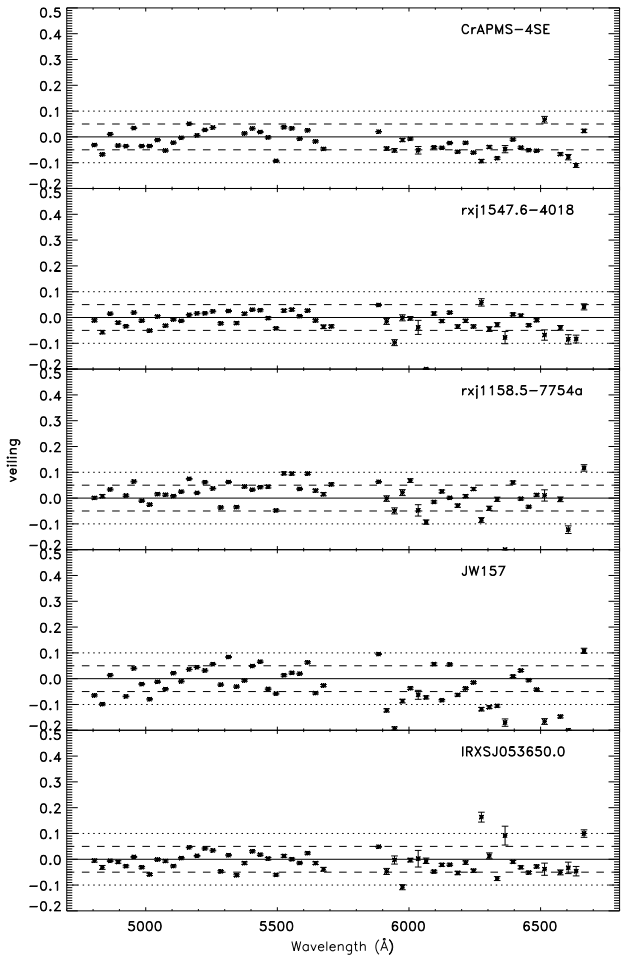}}
\caption{Dependence of veiling with wavelength for five stars in our sample 
with different temperatures. The dashed and dotted 
lines show, respectively, the veiling at levels 
$\pm$0.05 and $\pm$0.10. Negative veiling occurs when 
the photospheric lines in the reference star are shallower 
than in the target.} 
\label{fig:veiling}
\end{figure}

No clear trend in the veiling variation with wavelength 
was found for any of the stars. For all the objects, 
the veiling is centred near zero with some scatter 
(Fig.\,\ref{fig:veiling}). The values displayed in 
Table\,\ref{table:veiling} represent the mean veiling 
bluewards of 5700\r{A} and redwards of 5850\r{A}. The 
error bars represent the standard deviation in each 
region. The negative values and the highest error bars 
found in the red region are due to shallower 
photospheric lines and the presence of telluric lines 
in the object spectra, which are poorly fitted by 
the spectra of the rotationally broadened reference 
stars.

\section{Results and discussion}

\subsection{Systematic errors or an abundance dispersion inside the {\sc sfr}}
\label{sec:bias}

As can be seen from Tables\,\ref{table:parameters} and \ref{table:average}, 
the general dispersion observed in the [Fe/H] values of the stars in 
each {\sc sfr} is quite small. The same is true for 
the abundances of Si and Ni (Table\,\ref{table:elements} and 
Fig.\,\ref{fig:elements}). Possible exception is the Chamaeleon {\sc sfr}, 
where the dispersion of the measured metallicities is 
quite high for the 4 stars studied (although still of 
the order of magnitude of the abundance uncertainties).
Moreover, the abundances of the three species studied 
seem to be correlated, in the sense that higher iron 
abundances also correspond to higher [Si/H] and [Ni/H] 
values. 

A look at the temperature and metallicity values in Table\,\ref{table:parameters}
suggests, however, that a correlation may exist between metallicity and
effective temperature. Using all 19 stars in our sample, we find
a slope of $\sim$0.15\,dex/1000\,K for the points in a temperature-metallicity plot
(in this fit, [Fe/H]=0.0 would occur for T$_\mathrm{eff}\sim$5500\,K).
Although it may be unwise to use stars from different {\sc sfr}s to conclude,
this trend may suggest the existence of 
systematic errors in the derived stellar parameters. 

For a fixed metallicity, lower temperature stars tend to have deeper
spectral lines. The strongest of these may have extended damping wings. 
The fitting of gaussian profiles to measure the line-EWs 
could thus under-estimate the measurements, leading to lower
metallicity estimates. Although this hypothesis could in principle
explain the observed correlation, a look at individual lines in
our spectra (Fig.\,\ref{fig:spectrum}) shows that in the range of
EWs measured, the observed spectral lines are well approximated by gaussian
functions. This is actually expected, given that the large majority of the
lines used have EWs below 100-150\,m\AA. 

Furthermore, no correlation
is seen in the upper envelope of the temperature-metallicity plot of 
the large sample of stars presented in \citet[][]{Santos-2004b}, whose parameters were
obtained using the same line-lists and model atmospheres. 
We can add that the metallicities and efective temperatures
derived for main sequence dwarfs using the methodology presented here are in good agreement
with those found in the literature, derived using other methods
and model atmospheres \citep[e.g.][]{Santos-2004b,Casagrande-2006}.

\citet[][]{Morel-2004} studied the detailed effects of 
stellar spots affecting the determination of stellar 
parameters and metallicity for RS CVn stars. Their 
results show that the final abundances can be affected, 
although not always strongly, by the presence of 
photospheric spots. This issue may have particular 
importance when studying young, magnetically active 
stars like the ones in our sample. These effects 
possibly contribute to the observed metallicity-temperature
correlation.


We have also verified the possibility that a small 
residual veiling is causing the observed differences. 
For this purpose, we performed a simple test by 
increasing 3\% the measured {\sc ew} of iron lines 
used to derive stellar parameters for the two most 
metal deficient stars in the the Cha {\sc sfr} 
(RX J1140.3-8321, RX J1158.5-7754a). The value of
3\% is actually a conservative upper limit for the 
observed veiling errors (note that the uncertainties
listed in Table\,\ref{table:veiling} represent the observed
rms in the different veiling points, and not the error
on the average, which is much smaller). The resulting 
stellar parameters show only very small variations 
(below 10\,K in temperature and 0.01\,dex in 
surface gravity), and their iron abundances 
increased by a mere 0.02\,dex. Clearly, small 
amounts of un-accounted veiling cannot 
explain the observed abundance dispersion.

In any case, current data suggests that the observed
dispersions may be the result of some un-identified systematic effect
in our data, and not to the existence of a real dispersion in
the metallicities inside each {\sc sfr}. This conclusion also implies that the 
observed dispersions must be seen as upper limits.


\subsection{Other elements}

Chemical abundances of several elements are useful 
in distinguishing between different populations 
in the Galaxy \citep[e.g.][]{Nissen-2004,Fuhrmann-2004,Bensby-2005,Brewer-2006,Rich-2005,Fulbright-2005}. 
The observed differences are meant to reflect
different star formation histories or stellar initial mass functions 
in the different Galactic systems \citep[][]{Carigi-2006b}. It 
is thus interesting to check whether the abundances of the elements studied in 
the current paper for the six {\sc sfr}s are different 
from those found in field disk stars.

In Fig.\,\ref{fig:elements} we compare the abundance 
ratios [Si/Fe] and [Ni/Fe] listed in 
Table\,\ref{table:average} with those found for 
field planet-host stars and ``single'' stars from 
\citet[][]{Gilli-2006}. We remind the reader that 
Gilli et al. used the same procedure used in the 
current paper to derive the abundances of these 
elements (same model atmospheres, atomic line lists, 
{\em etc}), rendering such a comparison free from 
important systematic errors. 

\begin{figure}[t]
\resizebox{\hsize}{!}{\includegraphics{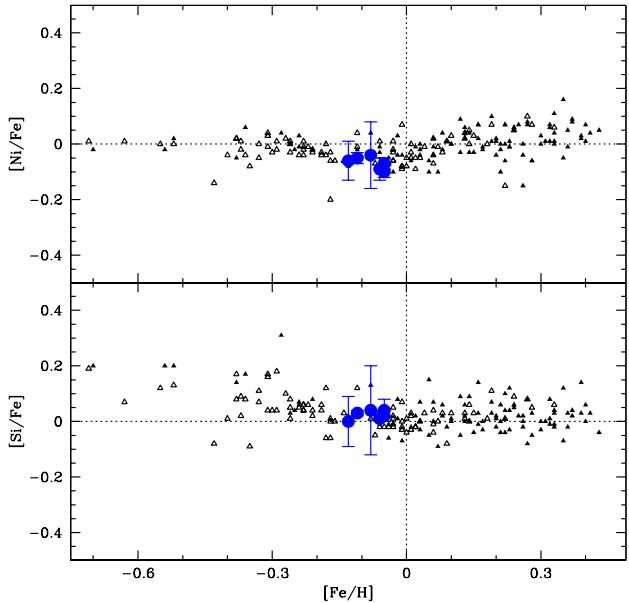}}
\caption{Abundance ratios [Si/Fe] and [Ni/Fe] as a function of [Fe/H] for the 
six {\sc sfr} (large circles) when compared with the 
same ratios for planet-host stars (filled triangles) 
and ``single'' stars (open triangles) from 
\citet[][]{Gilli-2006}. The dotted lines mark the 
solar abundances.}
\label{fig:elements}
\end{figure}

The results show that no clear differences exist 
between the abundance ratios of the studied {\sc sfr}s 
and those of field stars of the same metallicity. 
The chemical abundances in nearby {\sc sfr}s seem 
to reflect those of typical, thin disk stars.

\subsection{The metallicity distribution of nearby {\sc sfr}s}

In Table\,\ref{table:average}, we present the weighted 
average metallicities and abundance ratios for each 
{\sc sfr}, derived from the values listed in 
Table\,\ref{table:parameters}. In all cases the 
average abundances found are below solar. The 
metallicities for the six {\sc sfr}s vary from 
$-$0.13 ({\sc onc}) to $-$0.05 (Lupus and Taurus). 
Intriguingly, this result suggests that 
metal-rich star-forming clouds in the solar 
neighbourhood are rare. The implications of this
result are discussed in Sects.\,\ref{sec:subsolar} and \ref{sec:metalrich}.

In Fig.\,\ref{fig:histogram} we compare the metallicity 
distribution of our sample of six {\sc sfr}s (dashed 
line) with the same distribution for 1250 stars in 
the volume-limited {\sc coralie} planet-search sample 
\citep[see][]{Udry-2000,Santos-2004b}. The two 
distributions have similar average metallicities 
($-$0.08 for the {\sc sfr} sample, and $-$0.10 for 
the field stars). However, the metallicity distribution of 
the stars in the solar neighbourhood is clearly 
wider ($\sigma$=0.24\,dex), when compared with the 
one of the {\sc sfr}s ($\sigma$=0.033\,dex)\footnote{The 
average metallicity and dispersion of the field 
star sample is the same as the one found in the 
metallicity distribution for the subset of 94 
{\sc coralie} sample stars with high precision 
metallicity estimates from \citet[][]{Santos-2005a}.}.

To have some insight into the statistical meaning 
of the observed difference, we performed two different tests. 
First, a simple Kolmogorov-Smirnov test was done to 
verify the possibility that both samples belong to 
the same population. The result shows that 
P(KS)$\sim$14\%, a value that is clearly not 
significant, likely due to the small number of 
the observed {\sc sfr}s.

To better settle these issues we then performed 
a series of Monte-Carlo simulations. These were 
done to test the probability that in any given 
sample of randomly selected six metallicity points 
(taken from the field star sample) we would have 
a distribution similar to the one observed for the 
{\sc sfr}s. In the simulation, the metallicity values 
were computed taking into account the observed 
metallicity distribution for field stars presented 
in Fig.\,\ref{fig:histogram}. The methodology used is similar to 
the one presented in \citet[][]{Udry-2003}.


{
A total of 100\,000 data sets were simulated, 
and for each of them we computed the values for 
$\sigma$([Fe/H]) of the six 
simulated points. We then compared the number of 
cases where the simulated $\sigma$([Fe/H]) is smaller 
than the observed one (0.033\,dex).
}

\begin{figure}[t]
\resizebox{\hsize}{!}{\includegraphics{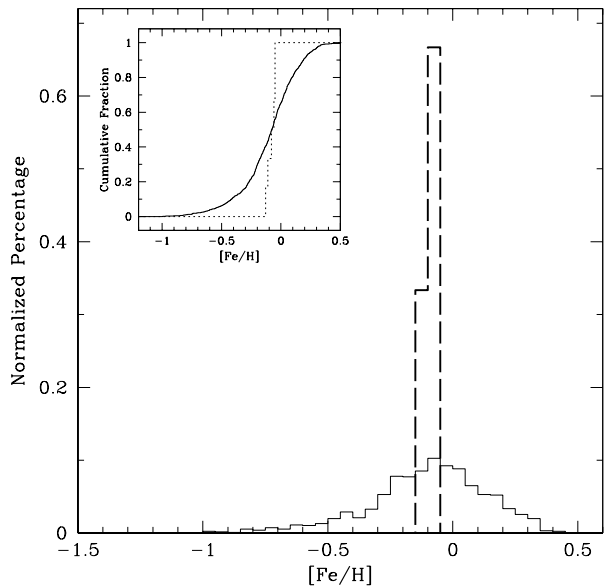}}
\caption{Metallicity distribution of solar-neighbourhood stars 
in a volume-limited sample 
(filled line -- see text) and of the metallicities of 
the six {\sc sfr}s analyzed in this paper (dashed line). 
In inset we show the cumulative functions of the two 
distributions.}
\label{fig:histogram}
\end{figure}

\begin{figure}[t]
\resizebox{\hsize}{!}{\includegraphics{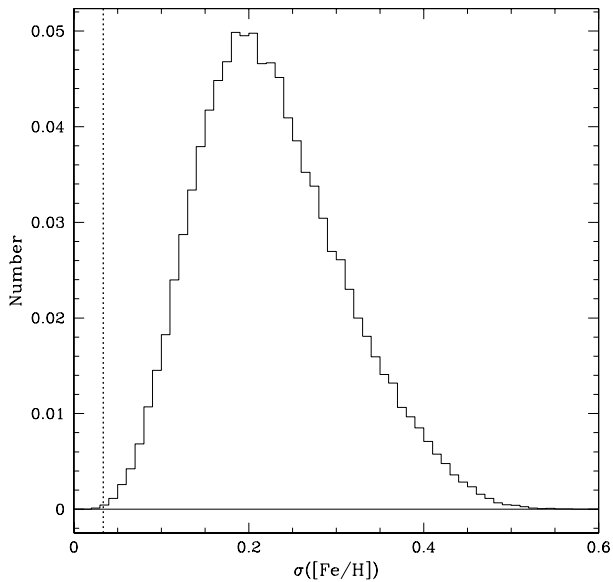}}
\caption{Distribution of 100\,000 simulated [Fe/H] dispersions ({\sc rms}). The dotted 
line denotes the observed {\sc rms} of the 6 star-forming regions. In only 0.02\% of 
the cases we obtained a dispersion below the observed value (0.033\,dex).}
\label{fig:simul}
\end{figure}



{
The results of these simulations show that the very small 
[Fe/H] dispersion observed is clearly 
at odds with the expected metallicity distribution 
of solar-type field stars. In our simulations, in 
only 0.02\% of the cases we obtained a value 
of $\sigma$([Fe/H]) lower than the observed 
0.033\,dex (Fig.\,\ref{fig:simul}). In other words, 
the very small dispersion observed for the 
metallicity distribution of the observed {\sc sfr}s 
is clearly significant.
}

\subsubsection{The small metallicity dispersion}

The small metallicity dispersion observed for the 
six star-forming regions (0.033\,dex) agrees with 
other recent studies done by \citet[][]{Nieva-2006}, 
who derived similar values for the dispersion of 
the abundances in nearby OB stars (0.04\,dex). 
When compared with local field dwarfs, a small 
dispersion is also observed in the metallicities 
of open star clusters \citep[][]{Twarog-1997}, nearby 
young FGK stars and nebular abundances 
\citep[see compilation by][]{Sofia-2001}.

Together with the literature results, the data 
presented in this paper suggests that the material 
in the nearby inter-stellar medium [{\sc ism}] is 
very uniform. In turn, this suggests that at 
any given Galacto-centric radius, the {\sc ism} 
is likely to be well mixed, and that mixing 
timescales are very short when compared with the 
time between different enrichment events
\citep[see discussion in][]{Roy-1995,Wielen-1996}. 
At a given epoch and Galacto-centric radius, 
the chemical abundances in the {\sc ism} are 
likely to be uniform.

If we consider that nearby {\sc sfr}s are good 
representatives of the current metallicity 
in the solar circle, the small scatter and 
slightly lower than solar abundances observed 
further imply that the metallicity distribution 
of main-sequence stars in the solar neighbourhood (with a much larger
metallicity dispersion) is strongly affected by stellar diffusion 
\citep[e.g.][]{Wielen-1996}. 

\subsubsection{The sub-solar metallicity of {\sc sfr}s}
\label{sec:subsolar}

Although the average metallicity of the observed {\sc sfr}s
is below solar, in Sect.\,\ref{sec:bias} we have seen
that some dependence seems to exist between the measured
metallicities and effective temperatures (a positive gradient
of $\sim$0.15\,dex/1000\,K is observed). Since our stars
are all cooler than the Sun (4664$<$T$_\mathrm{eff}<$5524\,K),
this could imply that some under-estimate of the metallicities
is done in our parameter estimate. We cannot exclude that the 
average metallicity of the observed {\sc sfr}s may be closer 
to solar, although we can reasonably exclude a value that is
strongly above.

Our results thus do not exclude (and somehow support) some models of Galacto-chemical evolution 
\citep[][]{Chiappini-2003,Carigi-2005} that predict that a slow chemical 
enrichment has happened in the Galactic disk during the last 4.5\,Gyr. 
We should add, however, that models of chemical evolution of the galaxy often
talk about oxygen as metallicity proxy, while
in the current paper we measured iron abundances. Contrarily to oxygen, iron is formed in
SN\,Ia events, that are supposed to continue even once the star-forming rate
has slowed down. It will thus be interesting to explore the impact of our
results on the models of galactic chemical evolution.

If, on the other hand, the sub-solar abundances observed are real (or
at least if the abundances are lower than those expected by models), 
the results could fit well a scenario where the Sun was 
formed in an inner region of the Milky-Way's 
disk \citep[later on experiencing orbital diffusion 
--][]{Wielen-1996,Edvardsson-1993a}, or that the 
nearby {\sc ism} suffered from a recent in-fall of 
metal-poor gas\footnote{Although this latter hypothesis seems unlikely given
the large quantities of printine gas needed \citep[see][]{Carigi-2006a}}. 
Sub-solar abundances were recently found for oxygen and 
carbon in B-type stars \citep[][]{Nieva-2006} and in the local bubble 
\citep[][]{Oliveira-2005}, although in the latter case 
dust depletion may account for the difference. 
These sub-solar metallicities are however at odds with some 
other studies of chemical abundances in \ion{H}{ii} 
regions \citep[][]{Esteban-2004,Esteban-2005}. 

Recent values for the solar oxygen abundance (lower than previously considered) 
also suggest that the {\sc ism} abundances may be closer to solar than previously thought \citep[e.g.][]{Chiappini-2003,Asplund-2004}. 
A recent review on this question is addressed 
by \citet[][]{Allende-2007}. However, if we consider that
heavy element settling has occurred in the Sun
\citep[][]{Lodders-2003,Bahcall-2006} decreasing its abundance
with respect to the initial value, the
lower-abundances problem can be even more important than
suggested by the mentioned values.

\subsubsection{On the origin of metal-rich stars}
\label{sec:metalrich}

As mentioned above, current models of galactic chemical
evolution predict that a slow chemical enrichment has happened
in the galactic disk during the past 4.5\,Gyr. Indeed,
the models cannot easily explain the existence of super metal-rich 
stars in the solar neighbourhood. 

Our results thus support the idea that super metal-rich stars (with 
metallicity clearly above solar) found in the solar 
vicinity may have been formed in inner Galactic disk 
regions. In a recent paper, \citet[][]{Ecuvillon-2007} 
present evidence that nearby metal-rich stars have 
kinematics similar to those of the Hyades (metal-rich) 
stream \citep[see e.g.][]{Famaey-2007}. This is also the 
case of stars with planetary-mass companions, that 
are found to be metal-rich when compared with single 
field stars \citep[][]{Santos-2004b,Ecuvillon-2007}. 
\citet[][]{Famaey-2007} and \citet[][]{Ecuvillon-2007} 
interpret these results on the basis of a scenario 
in which metal-rich stars in the solar neighbourhood 
(and in particular planet-hosts) could have formed 
in a more metal rich {\sc ism} and then suffered 
radial displacements due to a non-axisymmetric 
component of the Galactic potential, pushing them 
into the solar neighbourhood \citep[see also][]{Grenon-1999}. 

In this sense, it is interesting to see that for many chemical
elements the abundance ratios [X/Fe] seem to present a rising
trend as a function of [Fe/H] in the metal-rich domain, i.e., for [Fe/H]$>$0.0 
\citep[][]{Bensby-2005,Gilli-2006}.
Such trends seem very difficult to explain with current models
of galactic chemical evolution \citep[][]{Carigi-2006b}.
Curiously, for a given [Fe/H], higher abundance ratios are also observed for
stars in other galactic populations \citep[e.g.][]{Fuhrmann-2004,Zoccali-2006,Rich-2005}.

Also interestingly, there are suggestions that lithium abundances in 
planet-hosting, metal-rich stars show a systematic difference with respect to 
field stars in the temperature range between 5600 
and 5850\,K \citep[][]{Israelian-2004,Chen-2006}. We 
can speculate that the explanation for this difference 
can be related to a different lithium evolution in 
the inner Galactic disk.

\section{Concluding remarks}

In this paper, we present stellar metallicities for six 
nearby {\sc sfr}s, derived from high resolution spectroscopy 
of young {\sc wtts}s. The results show that all the 
studied {\sc sfr}s have metallicities between $-$0.13 
and $-$0.05\,dex, with a very small scatter of 
0.033\,dex around the mean value of $<$[Fe/H]$>$=$-$0.08\,dex
(compatible to solar within the uncertainties).

The analysis of the abundances of silicon (alpha-element) 
and nickel (iron-peak element) also suggests that the 
abundance ratios in the studied {\sc sfr}s are typical 
of those found in solar neighbourhood thin disk stars 
of similar metallicity. 

Together with other results from the literature, these 
observations suggest that the chemical abundances in 
the nearby {\sc ism} are extremely uniform and not strongly
above solar. Such a conclusion may have important implications 
for the models of Galactic evolution.

The results presented here have also important 
implications on planet searches around young 
(active) stars. Although these searches are severely 
limited in current radial-velocity surveys 
\citep[e.g.][]{Saar-1997,Santos-2000a,Paulson-2002}\footnote{See however recent
detection by \citet[][]{Setiawan-2008}}, 
several instruments are now being built that will 
change this situation. In particular, new generation 
adaptive optics systems like the {\sc eso sphere} planet 
finder or the Gemini {\sc gpi} project, as well as 
a new generation of interferometric instruments 
like {\sc prima} ({\sc eso}), will allow us to study, 
in unprecedented detail, the existence of long period 
planets around young solar-type stars. The rarity 
of metal-rich nearby star-forming regions may 
limit the goals of these projects if 
the metallicity-giant planet connection is still 
present for systems with long orbital periods.

In this context it will also be very interesting 
to study in detail the abundances of nearby young, 
post-T Tauri stars 
\citep[e.g.][]{Zuckermann-2004,Torres-2006}. Such 
studies would also help to settle the question 
about the metallicity of the local {\sc ism} 
\citep[e.g.][]{Sofia-2001}.

If stars with planets have originated from inner
galactic regions, it may be interesting to understand how
planetary evolution (and survival) may suffer with stellar 
orbital diffusion. This conclusion may further have
impact on the understanding of the frequency of planets
in the Galaxy.

As a very important by-product, the observations 
presented in this paper can further be used to 
investigate the effects of the metallicity on the 
properties of stellar populations such as rotation, 
multiplicity degree, or magnetic activity. These 
studies will be presented in a separate paper, 
James et al. (in prep.).

\begin{acknowledgements} 
We would like to thank M.F. Nieva, L. Carigi, and C. Chiappini for 
the fruitful discussions. We also thank the comments by the anonymous referee
who helped to improve the paper. NCS acknowledges the support 
from Funda\c{c}\~ao para a Ci\^encia e a Tecnologia (FCT), 
Portugal, in the form of a grant (references 
POCI/CTE-AST/56453/2004 and PPCDT/CTE-AST/56453/2004). This work was supported 
in part by the EC's FP6 and by FCT (with POCI2010 
and {\sc feder} funds), within the {\sc helas} international 
collaboration. JFG acknowleges the support from FCT 
(project PTDC/CTE-AST/65971/2006). We [DJJ] gratefully acknowledge financial 
support from the Programme National de Physique Stellaire 
de l'INSU for observations at OHP, and by support from 
an {\sc nsf} grant [AST-0349075] to Vanderbilt University.
\end{acknowledgements}

\bibliographystyle{aa}
\bibliography{santos_bibliography}

\end{document}